\documentstyle[12pt, epsf]{article}
\newcommand{\cf}{{\cal F}}
\newcommand{\cfbc}{{\cal F}_{B \leftrightarrow C}}

\newcommand{\lr}{\leftrightarrow}
\newcommand{\rl}{\leftrightarrow}

\newcommand{\st}{{s\bar{s}}}

\newcommand{\bp}{{\bf p}}

\newcommand{\btab}{\begin{tabbing}}
\newcommand{\etab}{\end{tabbing}}
\newcommand{\eqntimes}{\mbox{} \times}

\newcommand{\beqn}{\begin{equation}}
\newcommand{\eeqn}{\end{equation}}
\newcommand{\barr}[1]{\begin{array}{#1}}
\newcommand{\earr}{\end{array}}
\newcommand{\beqna}{\begin{eqnarray}}
\newcommand{\eeqna}{\end{eqnarray}}
\newcommand{\btablec}{\begin{table} \begin{center}}
\newcommand{\etablec}{\end{center} \end{table}}
\newcommand{\lapprox}{\stackrel{<}{\scriptstyle \sim}}

\newcommand{\gapproxeq}{\lower.7ex\hbox{$\;\stackrel{\textstyle>}
{\sim}\;$}}
\newcommand{\plabel}[1]{\label{#1}}
\newcommand{\pbibitem}[1]{\bibitem{#1}}
\input epsf

\begin{document}
\title{
\begin{flushright} 
\small{hep-ph/0007216} \\ 
\small{LA-UR-00-2143} 
\end{flushright} 
\vspace{0.6cm}  
\Large\bf Filter for strangeness in $J^{PC}$ exotic four--quark states}
\vskip 0.2 in
\author{Philip R. Page\thanks{\small \em E-mail:
prp@lanl.gov}\\{\small \em  Theoretical Division, Los Alamos
National Laboratory, Los Alamos, NM 87545, USA}}
\date{}
\maketitle
\begin{abstract}{Symmetrization selection rules for the decay of
four--quark states to two $J=0$ mesons are analysed in a non --
field theoretic context with
isospin symmetry. The OZI allowed decay of an isoscalar 
$J^{PC}=\{1,3,\ldots\}^{-+}$ exotic state to 
$\eta^{'}\eta$ or $f_0^{'}f_0$ is only allowed for four--quark 
components of the state containing one $s\bar{s}$ pair, providing a
filter for strangeness content in these states. 
Decays of four--quark $a_0$ states are narrower than otherwise expected.
If the experimentally
observed $1^{-+}$ enhancement in $\eta\pi$ is resonant, it is
qualitatively in agreement with being a four--quark state. 
}
\end{abstract}
\bigskip


PACS number(s): \hspace{.2cm}11.15.Pg \hspace{.2cm}12.38.Aw 
\hspace{.2cm} 12.39.Mk \hspace{.2cm} 13.25.Jx

\vspace{1cm}

Ever since the original work in the MIT bag model, 
it has been recognized
that multiquark states containing strange quarks can often have {\it lower}
energies than those with only
the equivalent light (up or down) quarks \cite{bag}, leading to the
prediction of the stability of strangelets.
For four--quark ($q\bar{q}q\bar{q}$)
states, the same conclusion
was reached in potential models \cite{semay}. 

In this Letter symmetrization selection rule II \cite{sel},
i.e. the case of isospin symmetry, is exhaustively 
analysed for the decay
of four--quark states to two $J=0$ hybrid or conventional
mesons  in QCD, expanding the earlier analysis 
\cite{sel}.
Decay topologies of (hybrid) 
mesons and glueballs to two (hybrid) mesons were considered
before \cite{sel}. The possibility of six--quark or higher multi--quark 
states is not considered.
It is shown that certain decays signal the presence of
strangeness in decaying 
$J^{PC}$ exotic four--quark states, providing an experimental tool to
verify the claimed presence of strangeness in these states. 
Decays also allow us to distinguish between the hybrid, glueball
 or four--quark
character of a decaying $J^{PC}$ exotic state. There are also implications
for non--exotic four--quark states.

We first consider states built only from isospin $\frac{1}{2}$ 
quarks, i.e. $u$ and $d$ quarks.
For four--quark states A  we are free to choose any 
basis to construct the flavour state. Labelling the quarks as $q_1\bar{q}_2q_3\bar{q}_4$,
and grouping $q_1\bar{q}_2$ and $q_3\bar{q}_4$ (denoted by $X$ and $Y$)
together, the four--quark flavour state is

\beqn \label{f1} |I_A I_A^z I_X I_Y \rangle \equiv
\sum_{I_X^zI_Y^z}
\;\langle I_A I_A^z | I_X I_X^z I_Y I_Y^z 
\rangle |X\rangle \; |Y\rangle  
\eeqn
where we summed over all isospin projections\footnote{
Because $X$ and $Y$ are merely labels, the states will be constructed 
to be representations of the label group, i.e. either symmetric or 
antisymmetric under $X\leftrightarrow Y$ exchange.
Models where the dynamics are truncated in such a way that 
$q_1\bar{q}_2$ occur in one meson, and $q_3\bar{q}_4$ in another, i.e.
where four--quark states are viewed as molecules of mesons, are {\it not}
included in our discussion. This is because, e.g. for
an $\eta\pi$ molecule, one can define $q_1$ and $\bar{q}_2$ to be in $\eta$. 
Label symmetry requires that $q_1$ and $\bar{q}_2$ can also be in 
$\pi$. But this is impossible by assumption. It should be noted that in QCD there is
nothing special about $q_1\bar{q}_2$ as opposed to $q_3\bar{q}_4$, so that
$X\leftrightarrow Y$ exchange is allowed.}. States can be verified to
satisfy the
orthonormality condition $\langle I_A I_A^z I_X I_Y |I_A^{'} I_A^{z'} I_X^{'} I_Y^{'} \rangle 
= \delta_{I_AI_A^{'}}\delta_{I_A^zI_A^{z'}}\delta_{I_XI_X^{'}}\delta_{I_YI_Y^{'}}$.

In this Letter we consider four--quark states with integral isospin.
When $I_A = 0$, the physical state is a linear combination of $|0\: 0\: 0 \: 0 \rangle$
and $|0\: 0\: 1 \: 1 \rangle$. For $I_A = 2$, the physical state is $|2\: I_A^z \: 1 \: 1 \rangle$.
Thus in both cases $I_X = I_Y$. When $I_A=1$, the physical state is a linear combination of 
$|1\: I_A^z 1\: 1 \rangle , \; |1\: I_A^z 1\: 0 \rangle$ and $|1\: I_A^z 0\: 1 \rangle$.
For $I_A=1$, we define new
states $|1\: I_A^z \pm\rangle \equiv \frac{1}{\sqrt{2}}(|1\: I_A^z 1\: 0 \rangle \pm |1\: I_A^z 0\: 1 \rangle)$.
The presence of $s\bar{s}$ pairs is now explored.
By convention, we choose a single strange
pair to correspond to labels $q_3 = s$ and $\bar{q}_4=\bar{s}$, so that
1 and 2 still labels $u,d$ quarks. The four--quark state is
$ |I_A I_A^z I_X s\bar{s} \rangle \equiv |X\rangle \; s\bar{s}$,
either isovector or isoscalar. 
Another possibility is $|00c\bar{c}s\bar{s}\rangle\equiv c\bar{c}
s\bar{s}$.
For two strange pairs, the state
is $|0 0 s\bar{s}s\bar{s} \rangle \equiv s\bar{s}s\bar{s}$.
Other states are obtained by freely interchanging strange, charm and
bottom quarks.
Explicit forms for some of the neutral states are given in Table 1.

\begin{table}[t]
\begin{center}
\begin{tabular}{|l|l||l|l|}
\hline 
\multicolumn{2}{|l||}{Isospin 2 four--quark:}&$|000\st\rangle$ & $\frac{1}{\sqrt{2}} (u\bar{u}+d\bar{d}) s\bar{s}$\\ \cline{1-2}
$|20 11\rangle$ &$\frac{1}{\sqrt{6}}(-u\bar{d}d\bar{u}-d\bar{u}u\bar{d}$&
& \\ \cline{3-4}
                &$+u\bar{u}u\bar{u}-u\bar{u}d\bar{d}-d\bar{d}u\bar{u}+d\bar{d}d\bar{d})$                           &\multicolumn{2}{|l|}{Isospin 1 four--quark:}\\ \cline{1-4}
\multicolumn{2}{|l||}{Isospin 0 four--quark:}&$|10 11\rangle$ & $\frac{1}{\sqrt{2}} (d\bar{u}u\bar{d}-u\bar{d}d\bar{u})$\\ \cline{1-2}
$|0000\rangle$  &$\frac{1}{2}(u\bar{u}u\bar{u}+u\bar{u}d\bar{d}+d\bar{d}u\bar{u}+d\bar{d}d\bar{d})$&$|10 + \rangle$ & $\frac{1}{\sqrt{2}} (u\bar{u}u\bar{u}-d\bar{d}d\bar{d})$\\
$|0011\rangle$  &$-\frac{1}{\sqrt{3}}(u\bar{d}d\bar{u}+d\bar{u}u\bar{d}$&$|10 - \rangle$ & $\frac{1}{\sqrt{2}} (u\bar{u}d\bar{d}-d\bar{d}u\bar{u})$\\ 
                &$+\frac{1}{2}(u\bar{u}u\bar{u}-u\bar{u}d\bar{d}-d\bar{d}u\bar{u}+d\bar{d}d\bar{d})$&$|10 1\st\rangle$ & $\frac{1}{\sqrt{2}} (u\bar{u}-d\bar{d}) s\bar{s}$\\ 
\hline 
\end{tabular}
\caption{Explicit neutral four--quark flavour states.
}
\end{center}
\end{table}

We shall be interested in decay and production $A\rl
BC$ processes in the rest frame of A. For simplicity we shall usually
refer to the decay process $A\rightarrow BC$, but the statements shall be equally
valid for the production process $A\leftarrow BC$.
The decay of an isospin $I_A$ four--quark state to two states with integral 
isospins $I_B$ and $I_C$ is considered \cite{note}.
The strong interactions include all interactions
described by QCD. The quarks and antiquarks in A are assumed to travel in all possible 
complicated paths going forward and backward in time and emitting and absorbing
gluons until they emerge in B and C. We shall restrict B and C to angular momentum $J=0$ states with valence quark--antiquark content and arbitrary gluonic excitation, i.e. to hybrid or conventional mesons. B and C can be radial
excitations or ground states, with $J^{P} =
0^{-}$ or $0^{+}$. If C--parity is a good quantum number, 
$J^{PC} = 0^{-+},0^{+-},0^{++}$ or $0^{--}$ are allowed. Since $0^{-+}$ and
$0^{++}$ ground state
meson states B and C are most likely to be allowed by phase space, they are used in
the examples.

Assume that states B and C are identical in all respects except, in principle,
their flavour and their equal but
opposite momenta $\bp$ and $ -\bp$.
Hence $B$ and $C$ have the same parity, $C$--parity, radial and gluonic 
excitation, as well
as the same internal structure. However, they are not required to have 
the same energies or masses \cite{sel}.
One possible example is $\eta$ and $\pi$.  

The decay  amplitude is a product of
the flavour overlap $\cf$ and the ``remaining'' overlap. We shall be interested
in the exchange properties of $\cf$ when the labels that specify the flavour
of the states $B$ and $C$ are formally exchanged, denoted by $B\rl C$. 
In cases where $\cf$ is non--zero and transforms into itself, which will be
of particular interest, define $\cfbc \equiv f \cf$. 

If a quark (or antiquark) in  A  ends up  
in the particle with momentum $\bp$, there is also the possibility that 
it would end up in the particle with momentum $-\bp$. 
Hence for a given topology in Figure 1,
e.g. 6a, there are in principle {\it two} topologically
distinct amplitudes. Furthermore, each of topologies 
4--6 is separately distinct. They are labelled analogous to
earlier conventions \cite{sel}.

\begin{figure}
\begin{center}
\leavevmode
\hspace{-.3cm}\hbox{\epsfxsize=5 in}
\epsfbox{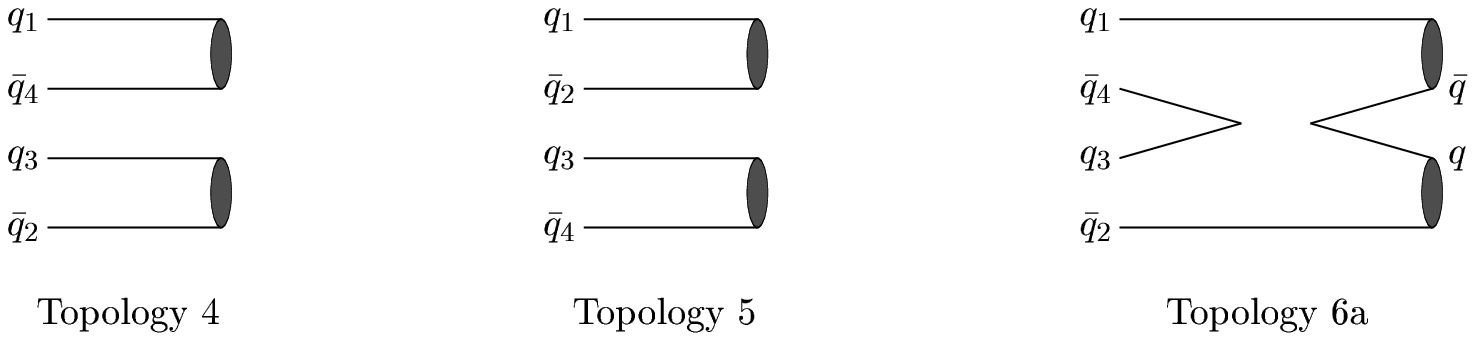}
\vspace{+.3cm}
\leavevmode
\hspace{-.3cm}\hbox{\epsfxsize=5 in}
\epsfbox{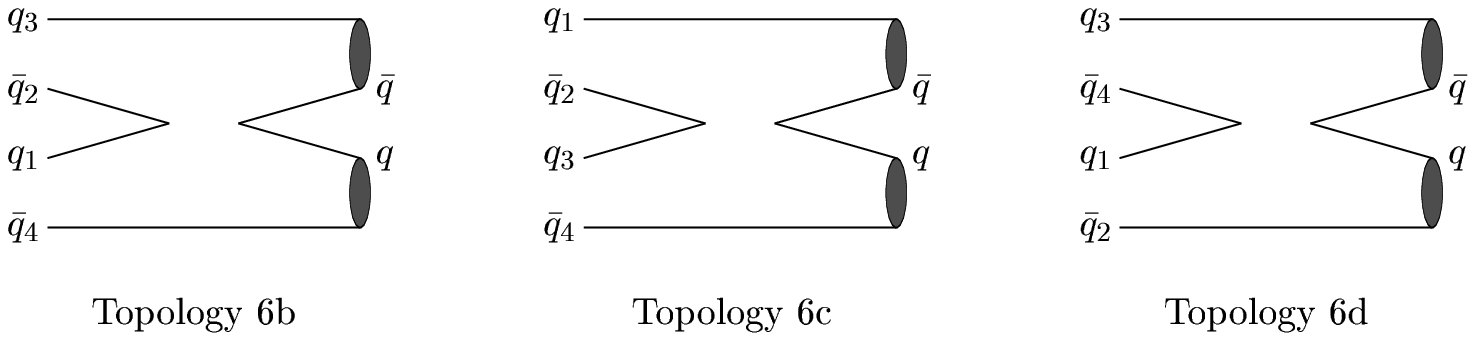}
\vspace{1.2cm}\caption{Connected topologies.}
\vspace{-1cm}
\leavevmode
\hspace{-.3cm}\hbox{\epsfxsize=5 in}
\epsfbox{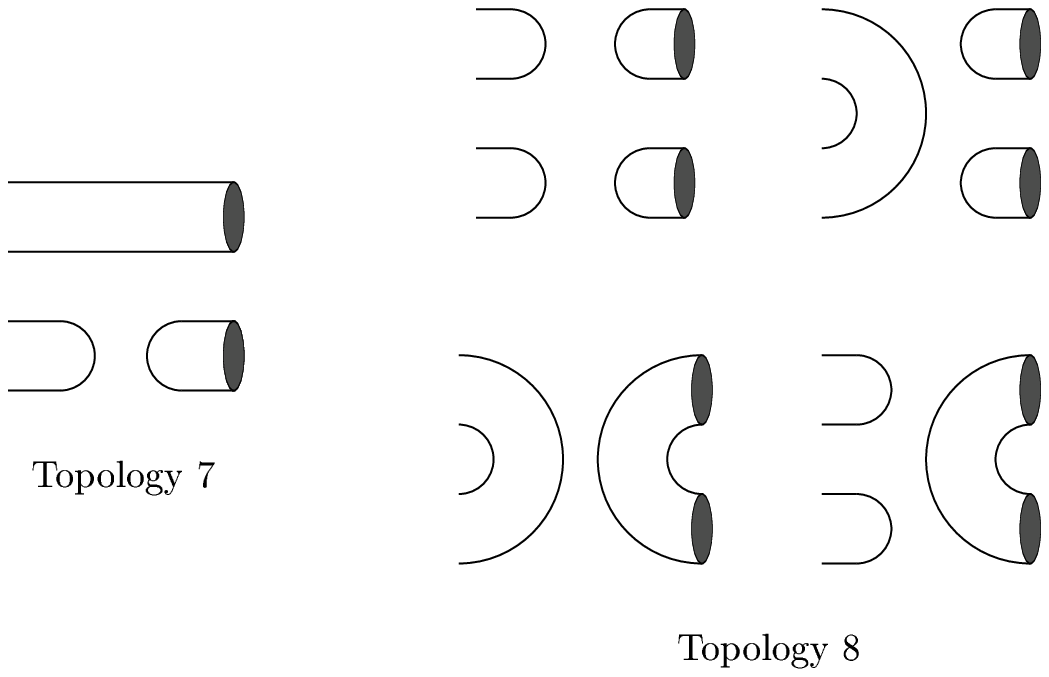}
\vspace{1.2cm}\caption{Disconnected topologies.}
\end{center}
\end{figure}

It is possible to omit the following proof of the results of this Letter
and continue directly to the statement of the results, which can be 
found where Table 2 is discussed in the text.

The flavour state of a $q\bar{q}$ pair is 

\beqn\plabel{fl}
|H\rangle = \sum_{h\bar{h}} H_{h\bar{h}} |h\rangle |\bar{h}\rangle
\hspace{1cm} \mbox{where} \hspace{.4cm} H_{h\bar{h}} = \langle I_H I_H^z | \frac{1}{2}h \frac{1}{2}-\bar{h}\rangle
(-1)^{\frac{1}{2}-\bar{h}}
\eeqn
and $|\frac{1}{2}\rangle = u,\; |-\frac{1}{2}\rangle = d,\; |\bar{\frac{1}{2}}\rangle = \bar{u}$ and $
|-\bar{\frac{1}{2}}\rangle = \bar{d}$. This just yields the usual $I=1$ flavour $-u\bar{d}, \;\frac{1}{\sqrt{2}}(u\bar{u}-d\bar{d}), \; d\bar{u}$ for $I^z = 1,0,-1$ and $\frac{1}{\sqrt{2}}(u\bar{u}+d\bar{d})$ for $I=0$. The advantage of this
way of identifying flavour is that any pair creation or annihilation 
that takes place will do so with $I=0$ pairs $\frac{1}{\sqrt{2}}(u\bar{u}+d\bar{d}) = \frac{1}{\sqrt{2}}\sum_{h\bar{h}}\delta_{h\bar{h}}|h\rangle |\bar{h}\rangle$ being formed
out of the vacuum, making the operator trivial.

In order to illuminate the method, we discuss the case where
only $u,d$ quarks participate in the decay. The presence of strange quarks
only simplifies the overlap. From Eqs. \ref{f1} and \ref{fl}, the
flavour overlap $\cf$ is

\beqna\plabel{ov}
\lefteqn{\hspace{-2.9cm}\sum_{ a_1\; a_2\; a_3\; a_4\; b\; \bar{b}\; c\; \bar{c}\; I_X^z\; I_Y^z } \;
\langle I_A I_A^z | I_X I_X^z I_Y I_Y^z \rangle\;
\langle I_XI_X^z | \frac{1}{2} a_1  \frac{1}{2} -a_2 \rangle\;
(-1)^{\frac{1}{2}-a_1}\; 
\langle I_YI_Y^z | \frac{1}{2} a_3  \frac{1}{2} -a_4 \rangle \; 
(-1)^{\frac{1}{2}-a_3}\; 
\nonumber } \\ & &   \eqntimes  
\langle I_B I_B^z | \frac{1}{2} b  \frac{1}{2} -\bar{b} \rangle\;
(-1)^{\frac{1}{2}-b} \;
\langle I_C I_C^z | \frac{1}{2} c  \frac{1}{2} -\bar{c} \rangle \;
(-1)^{\frac{1}{2}-c} \mbox{ KD} 
\eeqna
where ``KD'' is a set of Kronecker delta functions that specifies how the
quark lines connect in the
decay topology. Specialize to topology 6a as an example. From Figure 1
``KD'' is $\delta_{a_1 b} \delta_{a_2 \bar{c}} \delta_{a_3 a_4} 
\delta_{\bar{b} c}$. If one formally interchanges all labels $B$
and $C$ in Eq. \ref{ov}, it can be verified that
$\cf \rightarrow (-1)^{I_X+I_B+I_C} \cf$.
Since the overlap is non--zero only when $I_Y=0$ (due to the
$q_3\bar{q}_4$ pair annihilating), it follows by conservation of isospin
that
$I_A=I_X$, so that $\cf \rightarrow i \cf$, 
where $i\equiv (-1)^{I_A+I_B+I_C}$. Thus $f=i$.
This, as well as the fact that
the overlap vanishes when $I_Y=1$, are indicated in Table 2.

\begin{table}[ht]
\begin{center}
\begin{tabular}{|l|l|r||l|l|r||l|c|r|}
\hline 
\multicolumn{3}{|l||}{Isospin 0 four--quark}&
\multicolumn{3}{|l||}{Isospin 1 four--quark}&
\multicolumn{3}{|l|}{Isospin 2 four--quark}\\
\hline 
Top. & State & $f$ & Top. & State & $f$ & Top. & State & $f$ \\
\hline \hline 
4   & $|0000\rangle$ & $i$ & 4   & $|1I^z_A 11\rangle\;\dagger$ & $-i$ & 4   & $|2I^z_A 11\rangle$ & $i$ \\ 
    & $|0011\rangle$ & $i$ &    & $|1I^z_A + \rangle\;\dagger$ & $ i$ &  5   & $|2I^z_A 11\rangle$ & $i$ \\ \cline{7-9}
    & $|00\st\st\rangle\;\P$ & $i$ &    & $|1I^z_A - \rangle\;\S$ & $ i$  \\ 
5   & $|0000\rangle\;\P$ & $i$& 5   & $|1I^z_A 11\rangle\;\S$ & $ i$  \\ 
    & $|0011\rangle\;\S$ & $i$ & & $|1I^z_A + \rangle\;\dagger$ & $ i$  \\ 
    & $|000\st\rangle\;\P$ & $\ni$&    & $|1I^z_A - \rangle\;\dagger$ & $-i$  \\ 
    & $|c\bar{c}s\bar{s}\rangle\;\P$ & $\ni$ &    & $|1I^z_A 1\st\rangle\;\dagger$& $\ni$ \\ 
    & $|00\st\st\rangle\;\P$ & $i$ &  6a,b    & $|1I^z_A + \rangle$ & $ i$  \\ 
6a,b& $|0000\rangle$ & $i$&& $|1I^z_A - \rangle$ & $ i$  \\ 
    & $|000\st\rangle\;\ddagger$ & $i$&  & $|1I^z_A 1\st\rangle\;\amalg$& $ i$ \\ 
    & $|c\bar{c}s\bar{s}\rangle\;\P$ & $i$& 6c,d     & $|1I^z_A 11\rangle$ & $ i$  \\
    & $|00\st\st\rangle\;\P$ & $i$&  & $|1I^z_A + \rangle$ & $ i$  \\ \cline{4-6}
6c,d& $|0000\rangle$ & $i$\\
    & $|0011\rangle$ & $i$ \\
    & $|00\st\st\rangle\;\P$ & $i$ \\
\cline{1-3} 
\end{tabular}
\caption{Behaviour of the (non--vanishing) flavour overlap $\cf$ for the decay of the indicated four--quark state
to two mesons under $B\rl C$ exchange, i.e. $\cfbc = f \cf$, in the topology under consideration.
 The symbol $\ni$ denotes that 
$\cf$ has no simple transformation properties
under $B\rl C$ exchange, so that there is no symmetrization selection rule.
 If a state is not indicated for a given topology it means that $\cf$
vanishes.
When decay is not allowed by isospin conservation, ${\cal F}=0$ as expected.
This happens when
$\bf{I}_A \neq \bf{I}_B + \bf{I}_C$ or $I^z_A \neq I^z_B + I^z_C$,
or when $I_A=I_B=I_C = 1$ and $I^z_A = I^z_B = I^z_C = 0$.
$\dagger$ $\cf\neq 0$ only if $I_B\neq I_C$. 
$\S$ $\cf\neq 0$ only if $I_B = I_C=1$. 
$\P$ $\cf\neq 0$ only if $I_B = I_C=0$. 
$\ddagger$ In topology 6b ${\cal F} \neq 0$ only if $I_B = I_C=0$.
$\amalg$ $\cf\neq 0$ only in topology 6a.} 
\end{center}
\end{table}

Let $C^0_A$ be the C--parity of a neutral state $A$.  For charged
states (with no $C$--parity), we assume that at least one of the states in the isomultiplet
it belongs to has a well--defined C--parity, denoted by $C^0_A$.
G--parity conservation $G_A = G_B G_C$ and the relation
$G=(-1)^{I}C_A^0$ imply that $C^0_A = i$, as was noted in
section 2.2 of
ref. \cite{sel}. 

It was shown in Eq. 3 of ref. \cite{sel} that the
decay vanishes, called a symmetrization selection rule, if the parity $P_A = -f$.
If $f=i$, then $P_A = -f = -i = -C^0_A$, i.e. state A is $CP$ odd.
Since states B and C both have $J=0$, it follows by conservation of
angular momentum that an $L$--wave
decay would necessitate $J_{A} = L$. Hence states A have $J^{PC} = 
0^{+-}, 1^{-+}, 2^{+-}, 3^{-+}, \ldots$, which are all exotic $J^{PC}$
not found in the quark model, so that these states are not 
conventional mesons. A
charged state A (with no C--parity) should have a neutral isopartner with
the foregoing $J^{PC}$.
If $f=-i$, the same reasoning shows that states A have non--exotic $J^{PC} = 
0^{++}, 1^{--}, 2^{++}, 3^{--}, \ldots$.

The {\it results} of 
our analysis for topologies 4--6 are summarized in Table 2.
For topology 7 in Figure 2 
the flavour overlap has in general no simple transformation 
properties under $B \lr C$ exchange, corresponding to lack of 
symmetrization selection rules. Topology 8 is discussed further below. 
Topologies 4--6 are called ``connected'' and are allowed by the
Okubo--Zweig--Iizuka (OZI) rule \cite{ozi}, while topologies 7--8 are 
``disconnected'' and suppressed by the OZI rule.
{\it In the topology in Figure 1 under 
consideration an entry $i$ indicates that the decay
 of the corresponding four--quark component
vanishes for $J^{PC}=0^{+-},\; 1^{-+},\; 2^{+-},\; 3^{-+}, \ldots$
four--quark states. 
Ditto for an entry $-i$, except that
the four--quark state has 
$J^{PC}=0^{++ },\; 1^{--},\; 2^{++},\; 3^{--}, \ldots$.}
It immediately becomes clear that the decay of the four--quark states
with the $J^{PC}$ just mentioned is less than what one would na\"{\i}vely
expect, making them more stable.

To make the use of Table 2 clear, we consider the example to the decay
of an isovector $1^{-+}$ state to $\eta\pi$ in topologies 4--6. 
The $1^{-+}$ state is a linear combination of flavour wave functions
$|1I^z_A11\rangle$, $|1I^z_A+\rangle$, $|1I^z_A-\rangle$ and 
$|1I^z_A1s\bar{s}\rangle$. Referring to Table 2,
the $|1I^z_A11\rangle$ component decays in 
topology 4 only, $|1I^z_A-\rangle$ in topology 5 only and 
$|1I^z_A1s\bar{s}\rangle$ in topology 5 only. The $|1I^z_A+\rangle$
component does not decay.

The implications of Table 2 for the two $J^{PC}$ sequences
are now analysed.

{\it Decay of
$J^{PC}=0^{+-},\; 1^{-+},\; 2^{+-},\; 3^{-+}, \ldots$ four--quark states
to two $J=0$ mesons:}

We arrive at the following conclusions:

\begin{enumerate}

\item If $I_A=2$ or $I_A=I_B=I_C=1$, contributions from {\it all} four--quark
topologies vanish. They also vanish for all 
hybrid meson and glueball topologies
\cite{sel}. If $I_A=0$ and $I_B=I_C=1$, contributions from all connected
four--quark topologies vanish. They also vanish for the connected
hybrid meson topology \cite{sel}.

\item Contributions from all ``non -- fall apart'' connected topologies 6 vanish.

\item \plabel{zero} If $I_A=0$ and $I_B=I_C=0$, and the decay is non--vanishing, this comes from
either a single $s\bar{s}$ four--quark component
which decays via ``fall apart'' connected topology 5 or from disconnected 
topologies.
Also note that the decay cannot come from connected hybrid 
meson decay \cite{sel}.
Assuming the OZI rule that disconnected topologies are suppressed,
one discovers that a non--vanishing decay only comes from a 
single $s\bar{s}$ four--quark component. This isolates the presence of 
an $s\bar{s}$ component in the state, i.e. acts like a strangeness filter.
It has been noted \cite{lipkin1} that $u\bar{u},d\bar{d}$ components of a four--quark state
can in perturbation theory be expected to mix substantially via single gluon exchange with
$s\bar{s}$, although flavour mixing of this kind has been found to be $\lapprox 10\%$
in a model calculation \cite{semay}.

\item \plabel{one} If $I_A=1$ and $I_B\neq I_C$, decay does not come from the $|1I^z_A + \rangle$ component. 

\end{enumerate}

{\sf Examples: } There are no examples involving $\pi\pi$  final  states 
that are not
forbidden by well--known selection rules of QCD, e.g. G--parity or
$CP$ conservation,
or generalized Bose symmetry. Hence there is no new selection rules arising 
from item 1.
From the last two items we obtain the following examples:

Item \ref{zero}: Isoscalar $1^{-+},\; 3^{-+}, \ldots \rightarrow \eta^{'}\eta,
\; f_0^{'}f_0$ indicates a
four--quark component with a single $s\bar{s}$ in the initial state.

Item \ref{one}: Isovector $1^{-+},\; 3^{-+}, \ldots \rightarrow \eta\pi,\; \eta^{'}\pi,\;  f_0a_0,\;  f_0^{'}a_0$ does not come from a
$|1I^z_A + \rangle$ component in the initial state.

{\it Decay of
$J^{PC}=0^{++ },\; 1^{--},\; 2^{++},\; 3^{--}, \ldots$ four--quark states
to two $J=0$ mesons:}

In the cases that $I_A=1$ and $I_B\neq I_C$ some contributions vanish, making the states
narrower than otherwise expected.

{\sf Examples:} Isovector $0^{++ },\; 2^{++}, \ldots \rightarrow \eta\pi,\; \eta^{'}\pi,\;  f_0a_0,\;  f_0^{'}a_0$ is narrower than otherwise expected.

The decays can only be found to vanish by symmetrization selection rules 
if the quark structure of the decay is analysed. Models which only analyse
decay at the hadronic level, do not incorporate the selection rule:
The decay of four--quark $a_0(980)\rightarrow\eta\pi$ was recently
modelled at the hadronic level \cite{fari}.

The validity of the preceding discussion should be viewed within the
context of the restrictions on the final states $B$ and $C$ discussed
earlier.

This concludes the main results of this Letter.
A few final remarks are in order.

If one does not assume isospin symmetry
\cite{foot}, i.e. considers both QCD and QED,
the initial four--quark states
with different isospin will in general mix, yielding a complicated
behaviour for the flavour overlap under $B\rl C$ exchange in Table 2.
There are two exceptions. Firstly, for doubly charged states A 
(those with $I^z_A=\pm 2$ in the third column of
Table 1) $f=i$ in topologies 4 and 5.
Secondly, for all decays in topologies 6, $f=i$. Hence
 the symmetrization selection rule remains
valid in these cases even without isospin symmetry. One can verify that
each of these cases is an application of symmetrization selection rule I
of ref. \cite{sel}: the case without isospin symmetry. 

Consider topology 8 where two ``raindrops'' or a ``half--doughnut''
is created from the vacuum after the four--quark state has annihilated.
There are similar topologies for an initial meson or glueball \cite{sel}.
These topologies can be analysed without the need for isospin symmetry.
The ``half--doughnut'' can be shown to apply only for decays already known 
to vanish by $CP$ conservation or Bose symmetry \cite{sel}.
From the symmetrization selection rule III of ref. \cite{sel}, decay in
``raindrop'' topologies vanish in those cases where
the $B\rl C$ exchanged diagram is 
topologically distinct from the original diagram.

It needs to be emphasized that this Letter analyses the flavour structure
of various decay topologies in a generic way, which should subsume the
treatments of numerous models of QCD. However, it is {\it not} a 
field theoretic treatment, and can hence not be regarded as predictions
of QCD as a field theory. This becomes evident when one studies the following
condition for the validity of our conclusions. We assume that states B and C 
are identical in all respects except, in principle, their flavour.
Although this requirement is needed here, it is not sufficient, as a recent 
field theoretical analysis demonstrates \cite{large}: The requirement is not 
needed for at least on--shell $\eta$ and $\pi$ states $B$ and $C$ in 
a certain energy range and for certain quark masses.

A candidate state $\hat{\rho}(1405)$
with width $333\pm 50$ MeV, decaying to $\eta\pi$, and possibly to
$\eta^{'}\pi$, has been reported \cite{pdg}. 
It is interesting to note that a quark model calculation finds the
lightest  $1^{-+}$ four--quark state at $1418$ MeV, although it is an
isoscalar with flavour wave function $|000s\bar{s} \rangle$ \cite{semay}. 
The isovector state is heavier \cite{priv}.
If the $\hat{\rho}(1405)$  is resonant and has 
a substantial branching ratio of $\eta\pi$, 
this decay mode may discriminate against the hybrid interpretation
of the state. This is because  
only the (presumably suppressed) OZI forbidden hybrid meson topology 
contributes \cite{sel,large}.  
We predict that the OZI allowed decays of an isovector
 $1^{-+}$ only arise from certain four--quark components, so that the
detection of substantial branching ratios 
in $\eta\pi$ or $\eta^{'}\pi$ signals such a component. 

Useful discussions with L. Burakovsky and C. Coriano are acknowledged. This
research is supported by the Department of Energy under contract
W-7405-ENG-36.

\end{document}